\documentclass{aa}  
\usepackage{graphicx}
\usepackage{txfonts}
\usepackage{float}
\usepackage{xcolor}

\begin{document} 

  \title{The interacting nature of dwarf galaxies hosting superluminous supernovae}
  
  \author{Simon Vanggaard Ørum
          \inst{1}\fnmsep\thanks{Shared Lead Author}
          \and
          David Lykke Ivens
          \inst{1}\fnmsep$^\star$
          \and
          Patrick Strandberg
          \inst{1}\fnmsep$^\star$
          \and
          Giorgos Leloudas
          \inst{1}\fnmsep\thanks{\email{giorgos@space.dtu.dk}}
          \and
          Allison W. S. Man
          \inst{2}
          \and
          Steve Schulze
          \inst{3}
          }
  \institute{DTU Space, National Space Institute, Technical University of Denmark, Elektrovej 327, 2800 Kgs. Lyngby, Denmark 
         \and
             Dunlap Institute for Astronomy and Astrophysics, University of Toronto, 50 St George Street, Toronto ON, M5S 3H4, Canada
         \and
             Department of Particle Physics and Astrophysics, Weizmann Institute of Science, 234 Herzl Street, Rehovot 761000, Israel
             }
\abstract
   {Type I superluminous supernovae (SLSNe I) are rare, powerful explosions whose mechanism and progenitors remain elusive. Several studies have shown a preference for SLSNe I to occur in low-metallicity, actively star-forming dwarf galaxies.}
   {We investigate whether the host galaxies of SLSNe I show increased evidence for interaction. Galaxy interaction can trigger star formation and provide favourable conditions for these exceptional explosions to take place.}
   {Based on SLSN host galaxy images obtained with the \textit{Hubble Space Telescope} (\textit{HST}), we narrowed down  
   a sample  of 42 images obtained in the rest-frame ultraviolet over the redshift range between $0<z<2$.
   The number of host galaxy companions was measured by counting the number of objects detected within a given projected radius from the host. 
   As a comparison, we used two different Monte Carlo-based methods to estimate the expected average number of companion objects in the same \textit{HST} images, as well as a sample of 32 dwarf galaxies that have hosted long gamma-ray bursts (GRBs).  
   }   
   {About 50\% of SLSN I host galaxies have at least one major companion (within a flux ratio of 1:4) within 5~kpc. The average number of major companions per SLSN~I host galaxy is $0.70^{+0.19}_{-0.14}$. Our two Monte Carlo  comparison methods yield a lower number of companions for random objects of similar brightness in the same image or for the SLSN host after randomly redistributing  the sources in the same image. The Anderson-Darling test shows that this difference is statistically significant ($p$-value $<10^{-3}$) independent of the redshift range. The same is true for the projected distance distribution of the companions.  The SLSN I hosts are, thus, found in areas of their images, where the object number density is greater than average.
SLSN~I hosts have more companions than GRB hosts ($0.44^{+0.25}_{-0.13}$ companions per host distributed over 25\% of the hosts) but the difference is not statistically significant. The difference between their separations is, however, marginally significant with SLSN companions being closer, on average, than those of GRBs.}
   {The dwarf galaxies hosting SLSNe I are often part of interacting systems. This suggests that SLSNe I progenitors are formed after a recent burst of star formation. Low metallicity alone cannot explain this tendency.}
\maketitle

\section{Introduction}

More than a decade after their discovery, the mechanism that produces superluminous supernovae (SLSNe) remains unclear
\citep{2018SSRv..214...59M,2019ARA&A..57..305G,2019NatAs...3..697I}, particularly with respect to SLSNe I \citep{Quimby2011a}.
These powerful explosions still present an unsolved riddle despite the large samples collected to date \citep{Nicholl2015b,2018ApJ...860..100D,2018ApJ...855....2Q,2018ApJ...852...81L,AngusDES} and the various observational techniques available \citep[e.g.][]{Leloudas2015b,2018ApJ...864...45M,2019ApJ...876L..10E}. 

Because of the large distances to SLSNe I, a direct search for their progenitors is impossible.
A substantial effort has been undertaken to attempt to deduce the nature of SLSN progenitors indirectly, through observations of their host environments \cite[e.g.][]{Lunnan2014a,Leloudas2015a,Angus2016a,Perley2016a,Chen2016a,SUSHIESII}. 
Initially, it was suggested \citep{Lunnan2014a} that SLSN I hosts are similar to the hosts of long gamma-ray bursts (GRBs),  but \cite{Leloudas2015a} showed that they are even more extreme in terms of  mass, metallicity, and their specific star formation rate (SFR). 
The extreme nature of SLSN I hosts has now been confirmed with the largest available samples \citep{SUSHIESII}.
In addition, while there is broad agreement that SLSNe I preferentially occur in low-mass, star-forming dwarf galaxies, studies do not always agree on the physical reason behind the SLSN enhancement in these environments. While it is often attributed exclusively to low metallicity \citep[][]{Lunnan2014a,Perley2016a,Chen2016a}, other studies have proposed that a young progenitor age, deduced by the fraction of starbursting hosts and evidence for recent star formation, might be an equally important parameter  \citep{Leloudas2015a,Thoene2015a,SUSHIESII}.
These fundamental galaxy parameters are, of course, degenerate \citep[e.g.][]{Mannucci2010a} and  
\cite{2019arXiv191109112T} recently argued that it will not be an easy task to break the degeneracy with the current samples. 
Spatially resolved studies have only been possible for the most nearby events \citep{CikotaSLSNhosts} or with the Hubble Space Telescope  \citep[\textit{HST};][]{Lunnan2015a,Angus2016a}. 
Interestingly, \cite{Lunnan2015a} suggested that SLSNe I are less related to star formation within their hosts when compared to GRBs, but this result was not statistically significant.
In this paper, we study a different aspect of SLSN I hosts: in our studies \citep{Leloudas2015a,SUSHIESII}, we have often noticed that SLSN I hosts show signs of interaction or merging. This has also been noted by other authors for individual events \citep[e.g.][]{2017A&A...602A...9C}, but this effect has never been properly quantified.

The merger of gas-rich galaxies is capable of triggering star formation \citep{Schweizer1986,Barnes1991}.
When galaxies interact, their gas can lose angular momentum efficiently and collapse to form stars rapidly.
Systematic studies of the SFR of interacting galaxies versus isolated galaxies have indeed confirmed that interaction and mergers enhance the SFR of galaxies,
and the SFR enhancement is more pronounced for smaller separation between galaxy pairs \citep{Patton2011,Patton2013,Stierwalt2015}.
Major mergers, usually defined as having mass ratio between one  to four, are expected to cause the strongest starbursts \citep{Mihos1994,Kartaltepe2010}.
On the other hand, minor mergers, which have less extreme mass ratios of four to ten, could also boost star formation \citep{Kaviraj2011,Kaviraj2014} although their exact role hasn't been explored as much.
Existing studies of the occurrence of galaxy mergers are centred around the most luminous galaxies, that is, massive galaxies or those with substantially high SFR or black hole accretion activity \citep[e.g.,][]{Kocevski2012,Man2016, Cibinel2019}.
If the occurrence of superluminous supernovae (SLSNe) depends on the (specific) star formation rate of their host galaxies, 
we can expect the SLSNe to be preferentially found in interacting galaxies rather than isolated ones, independent of metallicity. 
The purpose of this work is to answer this question through a first statistical investigation of measuring the companion galaxy counts of SLSN galaxy hosts.
  
This paper is structured as follows: Section~\ref{sec:data} describes our sample. Section~\ref{sec:methods} outlines the methods used and Section~\ref{sec:results} presents our results. Implications and caveats are discussed in Section~\ref{sec:discussion} and our conclusions are summarised in Section~\ref{sec:conc}. 
The following  cosmological parameters were used throughout this study: $H_0 = 69.6$ \ km\ s$^{-1}$\ Mpc$^{-1}$, $\Omega_\text{M} = 0.286$ and $\Omega_{\Lambda} = 0.714$.

\section{Data sample}
\label{sec:data}

To examine the galaxy environment of SLSNe on the kpc scale, we used data obtained by \textit{HST}.
The high spatial resolution of \textit{HST} is required to obtain a clear view of galaxy hosts and to identify close companions that may appear to be blended in ground-based images. 
We searched through the Mikulski Archive for Space Telescopes (MAST)\footnote{\url{https://archive.stsci.edu/hst/}} and we retrieved images of 60 SLSN I host galaxies (observed sufficiently later than the SN explosion), obtained with various programmes. \footnote{Program numbers include: 12529 (PI: Soderberg), 12983 (PI: Yaron), 13022 (PI: Berger), 13025 (PI: Levan), 13326 (PI: Lunnan), 13858 (PI: De Cia), 14743 (PI: Nicholl), 14762 (PI: Maund), 15303 (PI: D'Andrea), 15140 (PI: Lunnan).}

\begin{figure}
  \resizebox{\hsize}{!}{\includegraphics{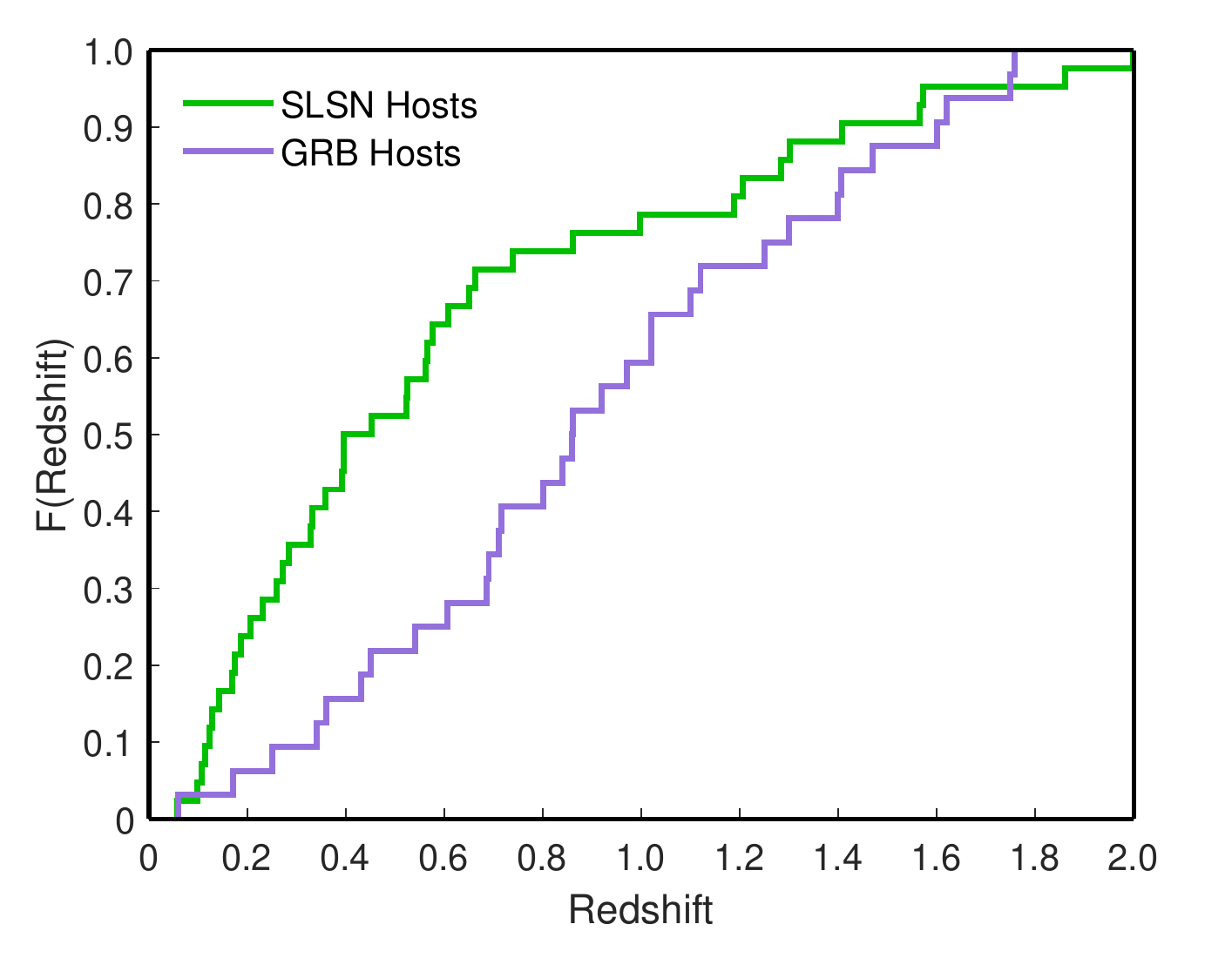}}
  \caption{Redshift distribution of the final SLSN I host sample (the complete sub-sample with $N = 42$) and the GRB host comparison sample. SLSN I hosts are skewed to lower redshifts than GRB hosts, with more than half of them found at $z < 0.5$.}
  \label{RedshiftDistribution}
\end{figure}

The SLSN redshift distribution (from $z = 0.057$ to $1.998$) is plotted in Fig.~\ref{RedshiftDistribution}.
As the SLSNe span a very wide redshift range, we must take care  to restrict the comparison to a consistent rest-frame wavelength.
Most galaxies have data in the rest-frame UV, while very few have rest-frame optical data. 
To maximise the sample size, we chose to work with UV data. 
The wavelength we probed is between $3000-3400$ \AA\ for most sources, however, for a few of them \citep[in particular the ones coming from DES;][]{AngusDES}, it can extend down to $2500$ \AA. 
We note that only ten galaxies from our final sample have \textit{HST} images in the rest-frame optical and their redshift distribution is very diverse (four of them are at $z > 1.2$).

We used the final drizzled images and, whenever available through the instrument pipeline, those that are also corrected for charge transfer efficiency, as retrieved by MAST. Although cosmic rays are normally removed during the \textit{HST} pipeline data processing, we noticed a handful of cases when the images were likely to have been severely contaminated by cosmic rays. This was particularly true for cases where only two raw images were combined to create the final reduced image. For a couple of cases, we were able to correct this effect by using custom reduced images that achieved a better result, such as for PTF11hrq \citep{CikotaSLSNhosts}.

The astrometry of the images was corrected by cross-checking and aligning  with sources from Gaia DR1 \citep{GaiaDR1}. In most cases, there were enough Gaia sources to produce a reliable solution and we found that the correction we had to apply was a minor one. In a handful of cases, however, there were fewer than three Gaia sources in the field and, in such cases, we relied on the native astrometry of the \textit{HST} images. 
We did not use SN images to align with the \textit{HST} host images (except in a couple of exceptions) as this would yield an astrometric precision beyond what is required for our project. Here, we are not interested in the exact location of the SLSN within its host at pixel precision \citep[e.g.][]{Fruchter2006,Lunnan2015a} but, rather,  the number of companions around the host within a radius of 5 kpc. So, as long as the host is correctly identified (see Sec.~\ref{sec:CompanionCounting}), the exact location of the SN is not of particular importance.

Several galaxies had to be removed from our sample. The most common reason was that the host galaxy could not be detected or identified with confidence in the images. We also had to remove the host of PTF10uhf, which is a large spiral galaxy, quite different from the general population of SLSN I hosts \citep{Perley2016a}. Our method would simply not work with this image, but this does not affect our study as it concerns the bulk population of dwarf hosts for SLSNe I. We also had to remove SN~1999as  because the UV image is too shallow to characterise the host system \citep[a star-forming region $>$10 kpc away from a larger host;][]{Angus2016a,SUSHIESII}.
Finally, iPTF13ehe had to be removed as the image contained significant spurious sources.
Table~\ref{rejected sample} summarises the objects that were removed from our initial sample and why.
Table~\ref{tab:UsedSample} contains all the objects in our final sample.

Since we are interested in the number of companions of the SLSN host, it is important to know whether such companions would be detected given the depth of the image. 
For this purpose, we investigated whether objects four times fainter than the SLSN host would yield a detection. This ratio was selected because it is often used to define major mergers in studies of interacting galaxies \citep[e.g.,][and references therein]{Man2016}. Usually, this definition refers to a mass ratio but in our case, where we work with monochromatic data, it refers to a flux ratio. 
We note that flux ratios in the UV can be decoupled from mass and this presents a limitation in our approach.
To quantify the effect of imaging depth on our analysis, we use the 
photometry of the SLSN host as well as the limiting magnitude obtained for the individual images.
If objects up to four times fainter than the SLSN host would yield a significant detection, we call these images `complete to major mergers'. This information is recorded in Tables~\ref{rejected sample} and \ref{tab:UsedSample}. We have achieved a good completeness in detecting major mergers within our sample (42/49 SLSN host images). 
Completeness to minor mergers, defined by a flux ratio of one to ten, is significantly reduced ($\sim$50\%). Minor mergers are not examined in this study. 
All results quoted in this paper (including figures) refer to the sub-sample complete to the detection of major mergers ($N = 42$), unless otherwise specified.

\begin{table}[tb]
    \caption[]{\textit{HST} SLSN I host sample and the objects removed.}
    \centering
    \begin{tabular}{ll}
    \hline\hline
    Total galaxies   & 60 \\ 
    \hline
    Undetected/uncertain host & SNLS07D2bv, PTF10vwg,  \\
       & PS1-10ky, PS1-10awh, \\
                    & PTF11rks, SN2011kf, \\
                    & iPTF13ajg, iPTF14tb \\
    Large spiral galaxy  & PTF10uhf \\
    Shallow UV image & SN1999as \\
    Many spurious sources & iPTF13ehe \\
    \hline
    \textit{Total} & 49 \\
    \hline
    Incomplete to major merger & DES15S2nr, DES16C3cv, \\ 
    detections & DES16C3ggu, iPTF13dcc, \\
                        & PTF09atu, PTF12gty, \\
                        & SCP06F6 \\
    \hline
    \textit{Complete sub-sample} & 42 \\
    \hline
    \end{tabular}
    \label{rejected sample}
\end{table}

\section{Methods}
\label{sec:methods}

\subsection{Counting companion galaxies} \label{sec:CompanionCounting}

Our purpose is essentially to determine whether SLSN I hosts are  parts of interacting systems. To this end, we created a MATLAB script that counts the number of companions for each SLSN host galaxy within a given radius.
We consider every individual source detected by the \texttt{SExtractor} software \citep{Bertin1996a} as a potential companion . 
The way \texttt{SExtractor} separates objects into multiple sources depending  on a number of input parameters. The values for these parameters were determined by experimenting on a few images and by visually inspecting  the result. In particular, we concluded that the following values were adequate for a few critical parameters: \texttt{DETECT}$\_$\texttt{MINAREA} $= 4.0$, \texttt{DETECT}$\_$\texttt{THRESH} $= 2.5$, \texttt{DEBLEND}$\_$\texttt{NTHRESH} $= 32$, \texttt{DEBLEND}$\_$\texttt{MINCONT} $= 0.01$ and \texttt{CLEAN}$\_$\texttt{PARAM} $= 3.0$.
For consistency, these values were used throughout our analysis. 
Therefore, we consider only sources detected at 2.5$\sigma$ above the background. 

We identify the host galaxy by using the SLSN coordinates reported in the literature\footnote{Collected in the webpage \url{https://slsn.info/} maintained by Ting-Wan Chen.} and by comparing with previous studies where the SLSN host has been identified \citep{Lunnan2015a,Angus2016a,Perley2016a,SUSHIESII,AngusDES} and cross-checking our identification with postage stamp images included in these publications. 

\begin{figure}
  \resizebox{\hsize}{!}{\includegraphics{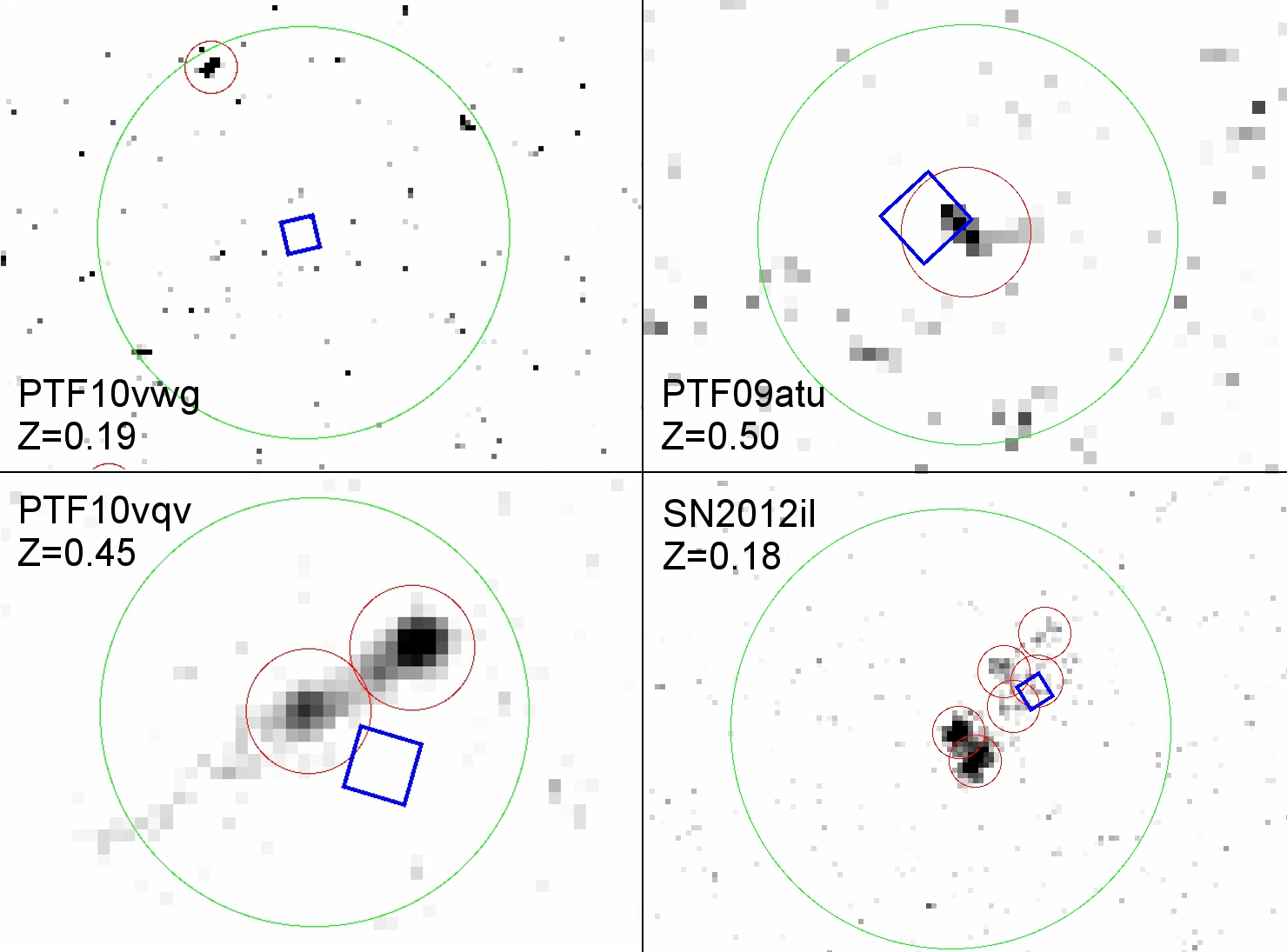}}
  \caption{Four representative example cases of SLSN host galaxies. Each red circle denotes a source detected as an individual object by \texttt{SExtractor}. The blue square marks the location of the SLSN. The green circle marks a radius of 5 kpc around the source that has been identified as the host galaxy (no reliable host has been identified in the upper left panel and the circle is centered on the SLSN).}
  \label{4Examples}
\end{figure}

Subsequently, we count the number of companions within a projected radius of 5 kpc around the host. 
Only companions within a flux ratio of 1:4 from the host (`major companions') are counted. Detections corresponding to minor mergers are ignored as they are not expected to significantly affect the evolution of the system.
Figure~\ref{4Examples} shows four representative examples. 

The upper left panel shows an example where the host galaxy cannot be securely detected (this source is therefore not included in our analysis). The upper right panel contains an example where a single galaxy (the host) is detected near the location of the SN explosion. The lower left panel shows a case where two sources are detected by \texttt{SExtractor} and the lower right panel a case with multiple objects within a radius of 5~kpc from the identified host.  

The radius of 5 kpc was selected because it corresponds roughly to the virial radius of dwarf galaxies, with masses similar to those of average SLSN I hosts \citep{SUSHIESII}. In addition, using a small radius minimises the contamination by chance projection.  

The number of major companion galaxies for every SLSN I with an identified host can be found in Table~\ref{tab:UsedSample}. In addition, we provide the distances of these companions as measured by the identified host galaxy (not the SN).

\subsection{Estimating the significance of companion counts by comparing with other sources in the images } \label{ComparisonSample}

In order to determine whether the SLSN hosts are found in regions where the object density is higher than average within the image the SLSN is found, we applied two statistical tests, based on the same \textit{HST} images and \texttt{SExtractor} catalogues as for the host detection. These tests compare the SLSN companion distribution to two distinct comparison samples generated through Monte Carlo simulations.

\begin{figure*}
\centering
   \includegraphics[width=15cm]{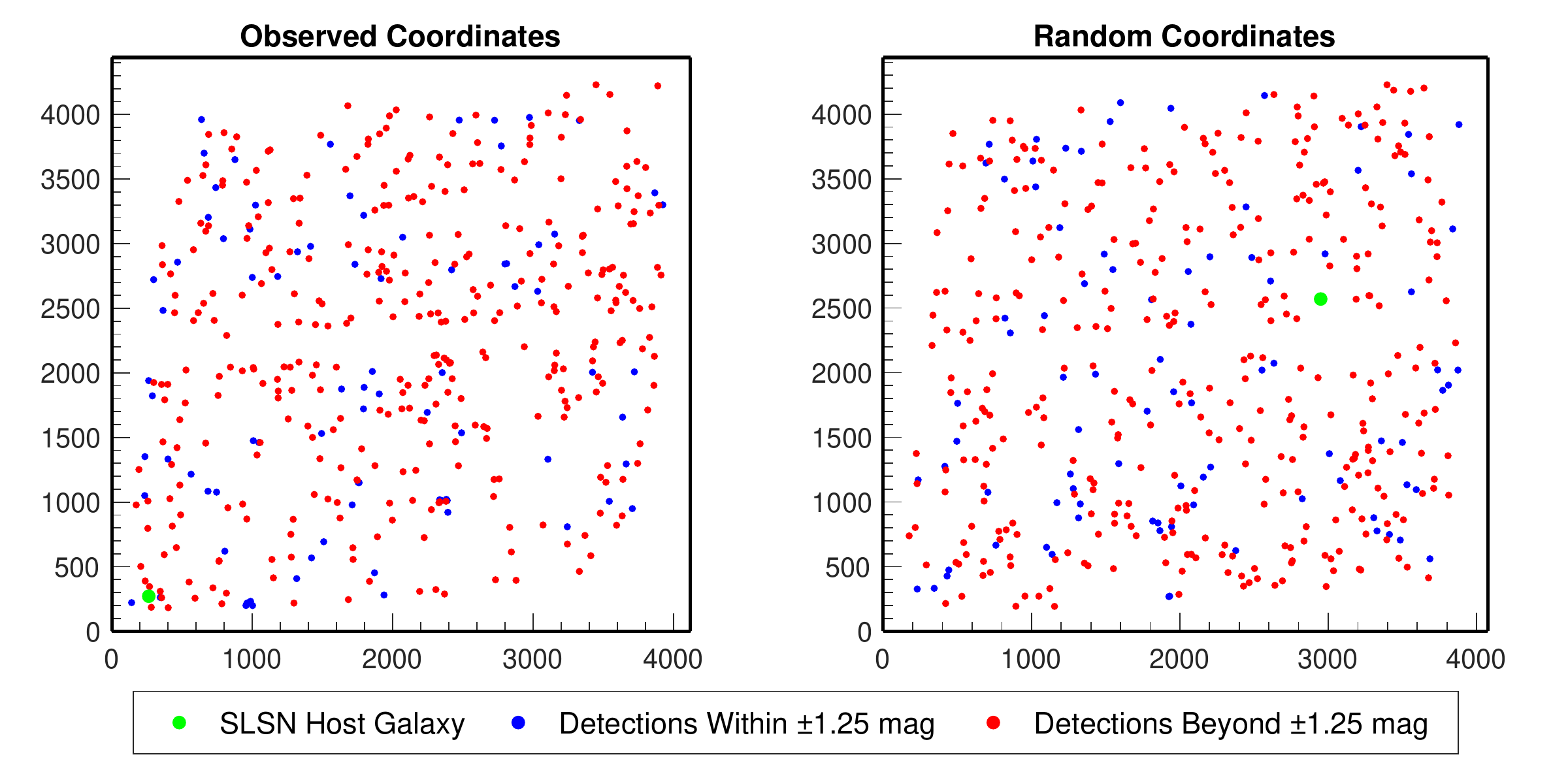}
     \caption{\textbf{Left}: Example map showing sources detected in the image of the host of PTF10aagc. 
     The SLSN host is shown in green, while objects within $\pm$1.25 mag from the host are shown in blue. All other sources are shown in red. 
     The masked regions appear clearly as white empty bands.
     In the \textit{random objects} test, only the companions of the blue sources are counted. 
     \textbf{Right}: Example map where the coordinates of all sources in the left panel have been re-generated randomly for an iteration of the \textit{random coordinates} test.}
    \label{fig:MapExample}
\end{figure*}

These methods are termed \textit{random objects} and \textit{random coordinates} and they are explained below. An example illustration can be found in Fig.~\ref{fig:MapExample}.
Similar Monte Carlo approaches are commonly employed to correct for chance alignment effects in the merger fraction of massive galaxies \citep{2011ApJ...738L..25W,2012ApJ...746..162N,Man2016}.

\subsubsection{Random objects} \label{RandomAnalysis}

With this method, we select a random object from the \texttt{SExtractor} catalogue and we count the number of its companions in a similar way as with the SLSN host. This procedure is repeated 1000 times and eventually we obtain the median of the given values as the best representation of the number of companions for objects in this image.
This procedure is repeated for every host galaxy image. 

However, not all random objects were allowed as we were not interested, per se, in the number of companions for objects of any brightness, but only those of a somewhat similar brightness to the host, to allow for a fair comparison. In addition, objects much fainter than the host would have completeness issues in their companion detections. It is for these reasons that we imposed a selection criterion that states that the random object should be within $\pm$1.25 magnitudes from the SLSN host (thus allowing the study of objects spanning a range of ten in flux). 

A limitation of this method is that the number of random objects available for selection may not be large enough. For most images, this is not a problem as there are a few hundred, or at least a few dozen, random objects with similar brightness to the SLSN host. 
However, for the PTF11hrq, PTF12dam, and SN2006oz hosts there are <10 random objects to select from. For this reason, we did not apply this method to these images.
To avoid many spurious detections that occur near the edges of the images and near the gap of the detector, these regions were masked out.

\subsubsection{Random coordinates}
\label{ScrambleAnalysis}

In the \textit{random objects} method, we used the exact coordinates as in the SLSN host image. To investigate further whether this was a particular (or special) object configuration in an image with the same object
density, we redistributed all objects in the image in a random way.
This simulation takes the \texttt{SExtractor} catalogue for a given SLSN host image and assigns randomly generated coordinates within the image boundary (excluding the masked regions) to all objects, including the host. We then count the number of companions around the new location of our host and repeat this process 1000 times.  Finally, we take the average of the number of companions for all realisations as the best representation for the number of companions for this host, if all image sources are randomly distributed.  
The \textit{random coordinates} method provides a comparison by generating a randomly distributed map, which helps us find out if the number of companions around the SLSN host galaxy is different than in a random distribution.\\

\subsection{Comparison with GRB host galaxies} \label{sec:GRBsample}

We also compared our SLSN I sample with a matched sample of long GRBs. Ideally, we would have liked to compare SLSN hosts to an unbiased and complete sample of dwarf galaxies at similar redshifts but, unfortunately, no such sample exists in the literature. 
GRB hosts are not an unbiased dwarf galaxy sample as GRBs also trace young star formation and favor low metallicity \cite[e.g.][]{Kruehler2015a}.
However, they have been extensively compared to SLSNe I in the past \citep[e.g.][]{Lunnan2014a,Lunnan2015a} and 
a relative comparison could  be very informative, particularly concerning the suggestion that SLSNe I might represent explosions of progenitor stars even younger than those of long GRBs \citep{Leloudas2015a,SUSHIESII}.

The GRB comparison sample has been selected in two different ways: for bursts before 2004 (the pre-\textit{Swift} era), we used the well-studied sample of \cite{Fruchter2006}, which consists of hosts observed mostly in rest-frame blue light. Reduced drizzled images were kindly provided to us by Andrew Levan in the form of cutouts around the location of the hosts. For bursts after 2004, we used a custom-made script that queried the \textit{HST} archive for all \textit{Swift} bursts that possessed at least one image at rest-frame $< 4000$ \AA, obtained at least 60 days after the GRB. In both cases, we only used GRBs at $z < 2$.

Our final sample consists of 32 long GRBs, summarised in Table~\ref{tab:UsedGRBSample}. We removed a couple of GRBs that otherwise fulfilled the above criteria, for reasons similar to our adjustments to the SLSN sample: GRB 171205A was removed because it was in a large spiral galaxy; GRB 090407 was removed because the host was not detected in the UV image; GRB 070125 was removed because the host could not be determined at all; and GRB 970828 was removed because the image was incomplete with regard to the detection of major mergers.
To identify the host galaxies, we were aided by previous studies  \citep{Fruchter2006,Blanchard2016a}. Similar to SLSNe, the exact explosion location and offset with respect to the host is not important for our study. 
The final redshift range for the GRB comparison sample is from $z = 0.059$ to $1.758$, similar to the SLSN one. However, the exact redshift distributions differ with the SLSNe being more skewed to lower redshifts (Figure~\ref{RedshiftDistribution}).
We note that the present GRB sample has been created to match our SLSN sample for comparison purposes and it is not meant to be unbiased in other respects \citep[such as e.g.][]{Hjorth2012a,Salvaterra2012a,Perley2016b}. A detailed study of the interacting nature of the full GRB host sample, including different rest-frame wavelengths, is beyond the scope of this paper.

\subsection{Statistical methods}
\label{subsec:Statisticalmethods}

To compare the distribution of companions for the SLSN~I hosts and the comparison samples, we used the Anderson-Darling K-sample test (AD test). This test is used to quantify if any differences between the respective distributions are statistically significant or not. 
This is both done for the total number of companions within 5 kpc as well as for their radial distribution. To estimate errors we bootstrap our samples from each group 10000 times, compute the mean for each iteration, and determine the 68.27\% confidence interval, with the upper and lower bounds used as our errors.

\subsection{Spatial resolution and binning}
\label{subsec:bin}

The spatial resolution of our images varies with redshift. As a result, the most nearby galaxies are better resolved than the ones at higher redshift, and there is a higher probability that they are separated into multiple sources by \texttt{SExtractor}. For this reason, we have performed an additional check to investigate this effect: images at $z < 0.3$ have been binned by $2 \times 2$ or $3 \times 3$ to `simulate a higher redshift'. After this binning, the resulting spatial resolution for all images at $z < 0.5$ is between $\sim0.2 - 0.3$ kpc/pix, with the exception of PTF11hrq and PTF10hgi (our two most nearby hosts below $z < 0.1$) which would require an even higher binning factor. 
The resolution difference effect is less pronounced between $0.5 < z < 2$, where all images have a spatial resolution between $\sim0.25 - 0.4$ kpc/pix.

\section{Results}
\label{sec:results}

We observe that about 50\% (20/42) of the SLSN I host galaxies have at least one major companion within 5~kpc. In addition, a few SLSN I hosts galaxies have multiple companions (Table~\ref{tab:UsedSample}).
The average number of companions per SLSN host galaxy is 0.70$^{+0.19}_{-0.14}$ (Table~\ref{tab:ResultAverageMedian}).
Here, the uncertainties have been determined by bootstraping 10000 samples and taking the 68.27\% confidence interval around the mean.
The respective number of companions obtained using our Monte Carlo comparison methods are $0.10^{+0.05}_{-0.05}$ (random objects) and $0.06^{+0.02}_{-0.01}$ (random coordinates).
The percentage of GRB hosts with major companions is $\sim$25\% (8/31) and the average number of major companions per GRB host is $0.44^{+0.25}_{-0.13}$.
A more detailed comparison of the sample distributions is given below. 

\subsection{Number of companions distribution}

\begin{figure}
  \resizebox{\hsize}{!}{\includegraphics{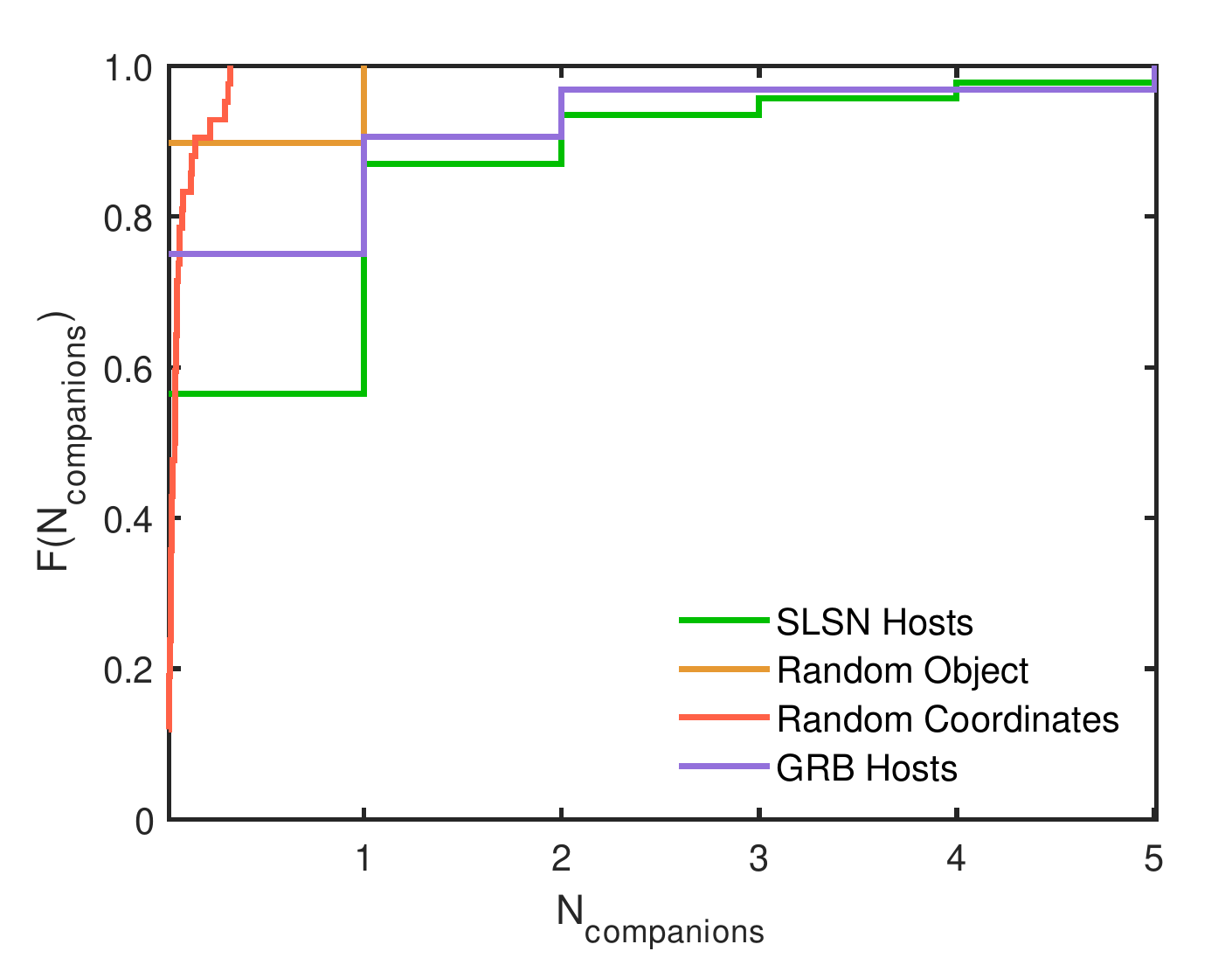}}
  \caption{Cumulative distributions showing the number of companions within 5~kpc for the SLSN host sample and the comparison samples for the full redshift range ($0 < z < 2$). The fractional number of companions in the \textit{random coordinates} sample is due to the Monte Carlo simulation and the very low number of companions per realisation.} 
  \label{fig:CumDens}
\end{figure}

Figure~\ref{fig:CumDens} shows the cumulative distributions for the SLSN host and the comparison samples for the number of companions found within 5~kpc.
An AD test yields that the difference between the SLSN host distribution and the Monte Carlo-generated distributions is statistically significant (see Table \ref{tab:ADtests}). 
On the contrary, the \textit{p}-value between the SLSN and the GRB distribution, considering all events at $z < 2$, is 0.07, which is less than 2$\sigma$. So, even if the GRB hosts appear to have fewer major companions than SLSNe, the difference is not statistically significant.

Comparing galaxy samples in such a wide redshift range ($0< z <2$) is problematic as it can include a number of potential biases, such as cosmic evolution (the mean properties of galaxies evolve with redshift) and Malmquist bias (only the brightest hosts are detected at high z). For this reason, we also examined different redshift bins, as in \cite{SUSHIESII}. Table \ref{tab:ADtests} contains the AD test results for these redshift bins. Only the lowest redshift bins ($z < 0.5$) have a large number of SLSN hosts ($N  > 20$). 
We observe that the SLSN hosts stand out from the Monte  Carlo-generated comparison samples and the differences are statistically significant (in terms of AD test \textit{p}-values) in every redshift bin. 
The comparison with GRB hosts yields higher \textit{p}-values, but the number of GRBs at the lowest redshift bins is very small ($N < 8$).
Table \ref{tab:ADtests} contains our reference results based on the complete sub-sample ($N = 42$). Use of the full sample ($N = 49$) yields consistent results.

\begin{table}[tb]
    \caption[Average values of counts]{Average number of major companions within 5 kpc.}
    \centering
    \resizebox{0.4\textwidth}{!}{%
    \renewcommand{\arraystretch}{1.25}
    \begin{tabular}{lc}
    \hline\hline
    Sample                  & Number of companions      \\\hline
    SLSN hosts              & $0.70^{+0.19}_{-0.14}$    \\ 
    Random objects          & $0.10^{+0.05}_{-0.05}$    \\
    Random coordinates      & $0.06^{+0.02}_{-0.01}$ \\
    GRB hosts & $0.44^{+0.25}_{-0.13}$ \\\hline 
    \end{tabular}}
    \label{tab:ResultAverageMedian}
\end{table}

\begin{table*}[tb]
    \caption[\textit{p}-value of Anderson-Darling test with varied conditions]{Anderson-Darling K-sample test \textit{p}-values comparing the SLSN host sample distribution to other distributions.}  
    \centering
    \begin{tabular}{lcllccc}
    \hline\hline
    z $^a$                      & Complet. $^b$              &  N$_{SLSN}$ $^c$       & N$_{GRB}$ $^d$         & SLSN - RO $^e$            & SLSN - RC $^f$             & SLSN - GRB $^g$ \\ \hline
    \multicolumn{7}{c}{Companion Number Distributions} \\ \hline
    0 -- 2.0                    &   yes                & 42                   & 32  & 1.89e-05 & 0.73e-03  & 7.06e-02\\
    0 -- 0.5                    &   yes                & 22                   & 7   & 1.06e-03 & 3.00e-13  & 7.98e-01\\
    0.1 -- 0.5 & yes & 20  & 6   & 4.73e-03 & 5.10e-09  & 6.82e-01\\
    0.5 -- 1.0 & yes & 11       & 12  & 1.98e-04 & 3.50e-02  & 8.43e-01\\
    0.5 -- 2.0 & yes & 20  & 25  & 1.00e-04 & 2.24e-02  & 2.53e-01\\
    \hline
    \multicolumn{7}{c}{Companion Distance Distributions}\\ \hline
    0 -- 2.0                    & yes                  & 42                   & 32  & 1.44e-07 & 5.61e-41  & 2.87e-02\\
    0 -- 0.5                    & yes                  & 22                   & 7   & 4.76e-05 & 8.09e-04  & 4.99e-01\\
    0.1 -- 0.5 & yes & 20  & 6   & 7.40e-05 & 8.09e-04  & 6.16e-03\\
    0.5 -- 1.0 & yes & 11  & 12  & 8.25e-02 & 8.25e-02  & 4.24e-01\\
    0.5 -- 2.0 & yes & 20  & 25  & 9.14e-05 & 9.14e-05  & 4.48e-01\\
    \hline\hline
    \end{tabular}\\
    \tablefoot{$^a$ Redshift bin. $^b$ Whether a completeness threshold was employed: `yes' indicates that only the SLSN sub-sample complete to the detection of major mergers was used. $^c$ Number of SLSN hosts per bin. $^d$ Number of GRB hosts per bin. $^e$ Comparison with the \textit{random objects} test. $^f$ Comparison with the \textit{random coordinates} test. $^g$ Comparison with the GRB sample.}  
    \label{tab:ADtests}
\end{table*}

\subsection{Companion distribution as a function of projected separation}
\label{subsec:radial}

In this section, we study in greater detail the radial distribution of the putative companions with respect to the host galaxies.
These were recorded in  Table~\ref{tab:UsedSample} for SLSN hosts and Table~\ref{tab:UsedGRBSample} for GRB hosts. 

\begin{figure}
  \resizebox{\hsize}{!}{\includegraphics{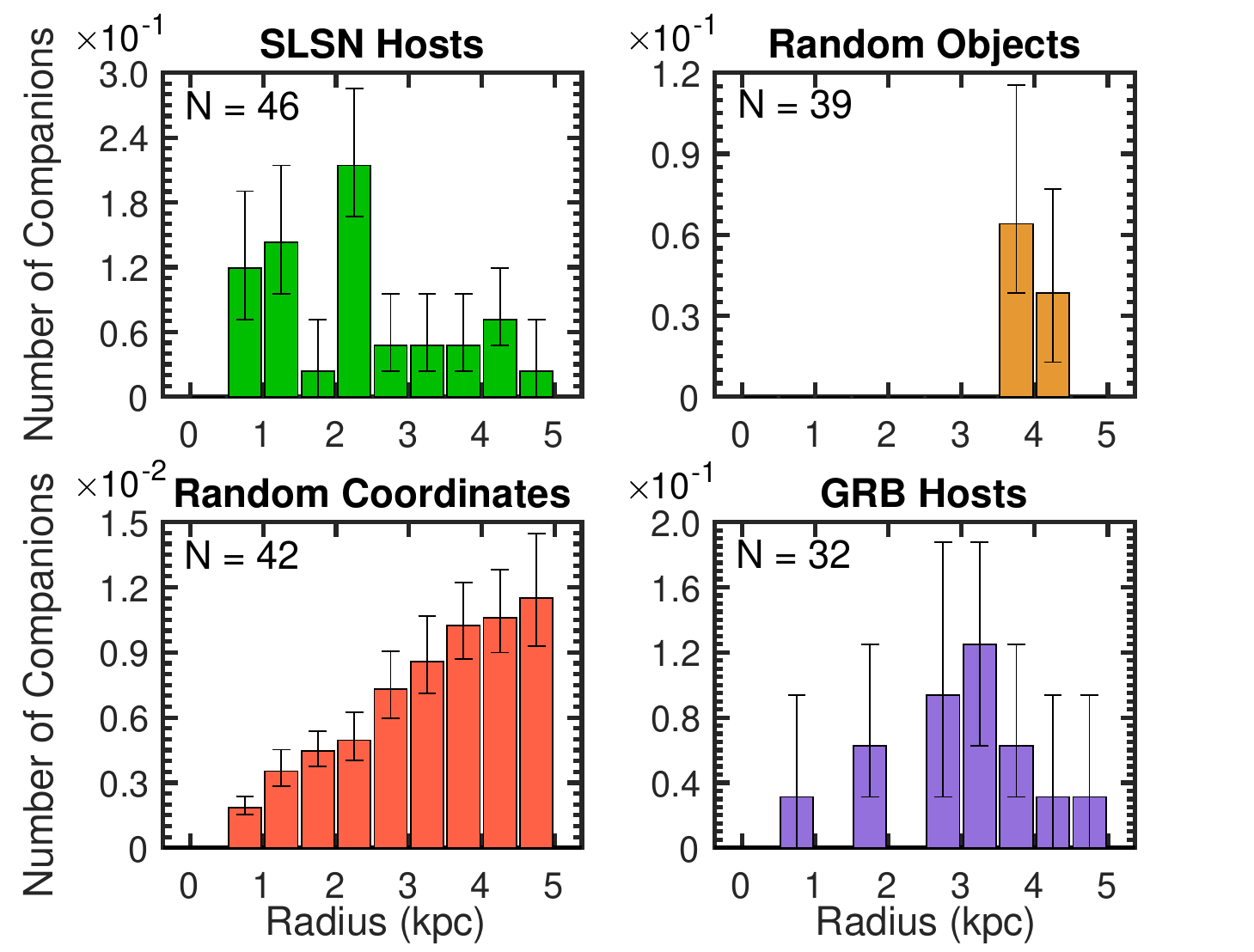}}
  \caption{Average number of major companions per host as a function of distance for SLSNe and comparison samples. The errors shown are given as a 68.27\% confidence interval from a 10000-sampled bootstrap. The SLSN companions are skewed to lower projected separations compared to the comparison samples. We note that the y-axis range differs across the panels.}
  \label{fig:AverageBins}
\end{figure}

\begin{figure}
  \resizebox{\hsize}{!}{\includegraphics{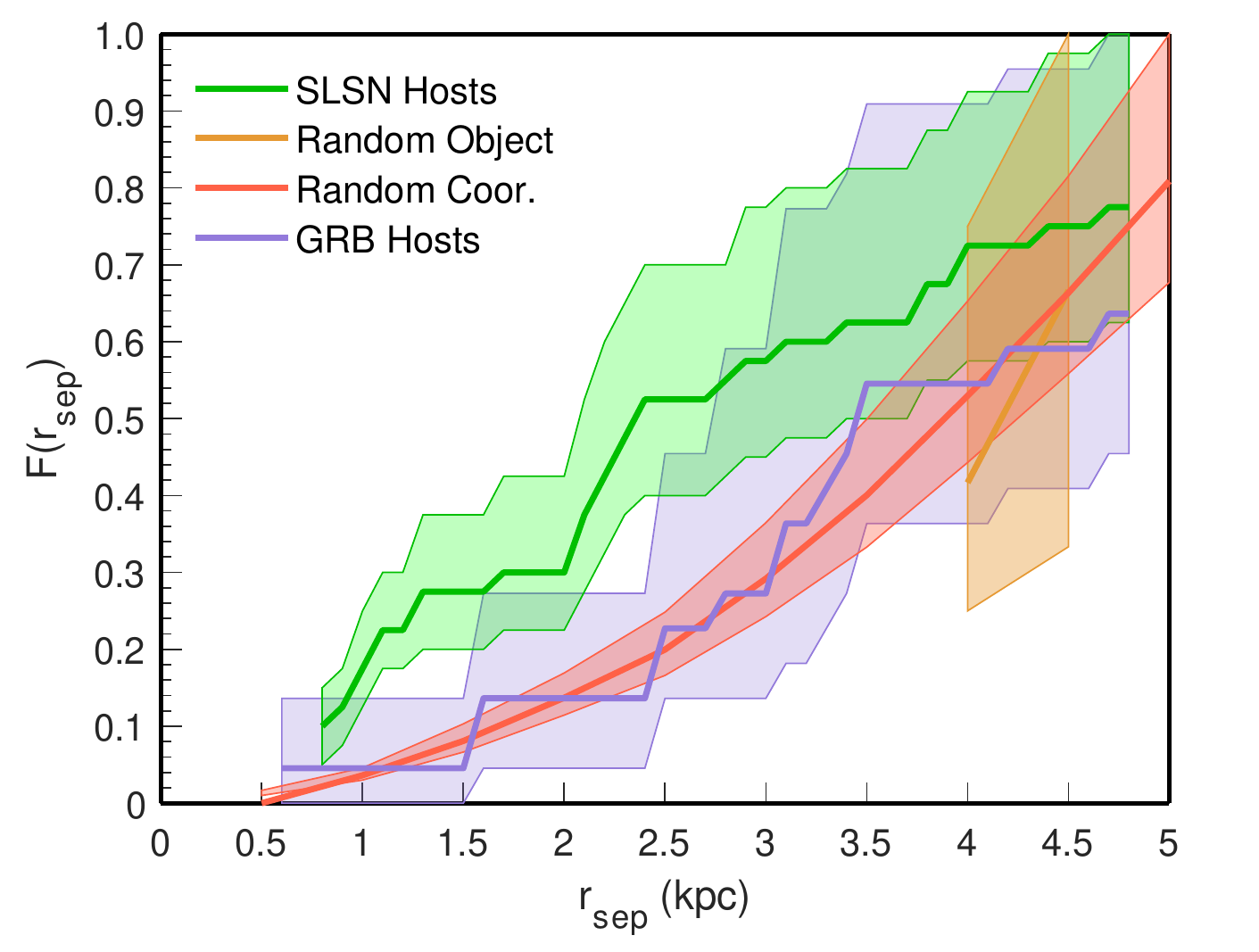}}
  \caption{Cumulative distribution of companions as a function of distance. The errors shown are given at 1$\sigma$ confidence intervals from bootstrapping 10000 samples. The two Monte Carlo-generated samples are plotted with a resolution of 0.5 kpc for visualisation purposes.}
  \label{fig:CumDist}
\end{figure}

We created equal projected separation bins of 0.5~kpc and determined the overall average number of companions for the SLSN hosts. 
The corresponding histogram showing the radial distribution of companions around SLSN hosts is shown in Fig.~\ref{fig:AverageBins} (upper left panel). 
Similarly, the other histograms show the corresponding distributions for our comparison samples. The errors for the simulated samples are computed the same way as for the general SLSN sample (see Section \ref{subsec:Statisticalmethods}). 
Figure~\ref{fig:CumDist} shows the corresponding cumulative distribution functions.
The second part of Table~\ref{tab:ADtests} contains AD tests between the radial distributions at a resolution of 0.5 kpc. It is possible to confirm that the differences between the SLSN host distribution and the  two Monte Carlo-generated distributions are statistically significant in every redshift bin. 

In addition, the difference between SLSNe and GRBs is marginally statistically significant (between 2 and 3$\sigma)$ both for the full redshift range and the $z = 0.1 - 0.5$ redshift bin (modulo the low number of GRBs in this bin). This means that the projected separation between the hosts and their major companions are smaller on average for SLSN hosts than for GRB hosts. This difference is also visible in Fig.~\ref{fig:CumDist}.
This is very important both because the nearest companions have a higher probability of being real companions and because the probability and impact of interaction will be higher on the host.

\subsection{Effects of redshift and resolution}
\label{subsec:redres}

As described in Sec.~\ref{subsec:bin}, we binned SLSN host images at $z < 0.3$ to investigate  the effect of resolution in our ability to detect companion galaxies as separate  objects. This redshift range was selected because this is where the variations in spatial resolution were more pronounced and where a number of SLSN hosts were found to have multiple companions.

This experiment revealed that a number of companions were no longer detected as individual objects. The SLSN hosts affected were: PTF09cnd, PTF10hgi (where the number of companions was reduced from one to zero), PTF10aagc, PTF10bfz (from two to one), and SN 2012il (from five to one). 
Overall, the number of major companions in our SLSN sample was reduced after degrading the resolution of our images. 
This indicates that resolution does play a vital role in uncovering interacting companions for dwarf galaxies and that this is done easier at low redshift.  
This effect might be enhanced by our choice of wavelength as galaxies appear more patchy in the UV than in the optical, due to the fact that UV light traces star-forming regions.
Nonetheless, the number of SLSN hosts with at least one major companion remains high (18/42 or $>$40\%) and all differences between the SLSN companion distribution and the Monte Carlo-generated comparison samples (re-computed on the binned images) remain statistically significant.

Investigating the difference with GRBs is more complex precisely due to the different redshift distribution. However, the GRB sample is subject to the same limitation and GRB 100316D stands out as the only GRB host with five detected companions in our method. We note that this is a well-known complex system at $z = 0.059$, which also stands out as a low-luminosity GRB, different than the rest of the sample. \citep{2011MNRAS.411.2792S,2017MNRAS.472.4480I}.
Repeating our binning experiment for GRBs (to an approximate spatial resolution of 0.35 kpc/pix) and comparing with the (binned) SLSN distribution, we notice no significant differences in the comparison between these distributions.

We note, however, that the very low-redshift events PTF10hgi and, especially PTF11hrq and GRB 100316D, still have a higher resolution because a binning higher than $3 \times 3$ (and up to $9 \times 9$) would be required. 
We did not want to modify our images to such a degree.
An alternative way to obtain a more fair comparison between SLSN and GRB hosts is to completely remove the very-low-redshift objects. For this reason, in Table~\ref{tab:ADtests} we also provide results for a redshift bin $0.1 < z < 0.5$.

\begin{table*}[tb]
\caption[Used sample]{SLSN I host galaxies used in our analysis.}
\label{tab:UsedSample}
    \centering
    \begin{tabular}{lrrrrr}
    \hline\hline
    Host galaxy & Redshift  & Refs. $^a$ & Complet. $^b$ &  N $^c$     & Companion distance \\
                &           &   &       &       & (kpc)\\
    \hline
    DES13S2cmm  & 0.663     & (1)  & yes   & 1     & 4.1 \\
    DES14C1fi   & 1.302     & (1)  & yes   & 3     & 3.2, 3.9, 4.1 \\
    DES14X3taz  & 0.608     & (1)  & yes   & 0     & - \\
    DES15C3hav  & 0.392     & (1)  & yes   & 0     & - \\
    DES15E2mlf  & 1.861     & (1)  & yes   & 0     & - \\
    DES15S2nr   & 0.220     & (1)  & no    & 0     & - \\
    DES15X1noe  & 1.188     & (1)  & yes   & 0     & - \\
    DES15S1nog  & 0.565     & (1)  & yes   & 1     & 2.2 \\
    DES15X3hm   & 0.860     & (1)  & yes   & 0     & - \\
    DES16C2nm   & 1.998     & (1)  & yes   & 0     & - \\
    DES16C3cv   & 0.727     & (1)  & no    & 1     & 0.9 \\
    DES16C3dmp  & 0.562     & (1)  & yes   & 0     & - \\
    DES16C3ggu  & 0.949     & (1)  & no    & 0     & - \\
    iPTF13bjz   & 0.271     & (2)  & yes   & 0     & - \\
    iPTF13cjq   & 0.396     & (2)  & yes   & 1     & 2.4 \\
    iPTF13dcc   & 0.431     & (3)  & no    & 0     & - \\
    iPTF14dck   & 0.576     & (4)  & yes   & 0     & - \\
    iPTF14dek   & 0.332     & (4)  & yes   & 2     & 2.2, 4.8 \\
    PS1-10bzj   & 0.650     & (5)  & yes   & 0     & - \\
    PS1-10pm    & 1.206     & (5)  & yes   & 1     & 3.5 \\
    PS1-11afv   & 1.407     & (5)  & yes   & 1     & 2.4 \\
    PS1-11aib   & 0.997     & (5)  & yes   & 0     & - \\
    PS1-11ap    & 0.524     & (5)  & yes   & 0     & - \\
    PS1-11bam   & 1.565     & (5)  & yes   & 0     & - \\
    PS1-11bdn   & 0.738     & (5)  & yes   & 0     & - \\
    PS1-11tt    & 1.283     & (5)  & yes   & 1     & 2.3 \\
    PS1-12bmy   & 1.572     & (5)  & yes   & 1     & 2.5 \\
    PS1-12bqf   & 0.522     & (5)  & yes   & 0     & - \\
    PTF09as     & 0.186     & (6)  & yes   & 0     & - \\
    PTF09atu    & 0.501     & (6,7)  & no    & 0     & - \\
    PTF09cnd    & 0.259     & (6,7)  & yes   & 1     & 1.2 \\
    PTF10aagc   & 0.207     & (6)  & yes   & 2     & 1.0, 1.4 \\
    PTF10bfz    & 0.169     & (6)  & yes   & 2     & 0.9, 2.3 \\
    PTF10bjp    & 0.359     & (6)  & yes   & 0     & - \\
    PTF10hgi    & 0.098     & (6)  & yes   & 1     & 0.8 \\
    PTF10nmn    & 0.124     & (6)  & yes   & 0     & - \\
    PTF10vqv    & 0.452     & (6)  & yes   & 1     & 2.9 \\
    PTF11hrq    & 0.057     & (6,8)  & yes   & 1     & 0.8 \\
    PTF12dam    & 0.108     & (8,9) & yes   & 1     & 1.1 \\
    PTF12gty    & 0.177     & (6)  & no    & 1     & 3.4 \\
    PTF12mxx    & 0.327     & (6)  & yes   & 4     & 1.1, 2.5, 3.9, 4.5 \\
    SCP06F6     & 1.189     & (5,7)  & no    & 0     & - \\
    SN2005ap    & 0.283     & (7)  & yes   & 1     & 1.2\\
    SN2006oz    & 0.396     & (10,11)  & yes   & 0     & - \\
    SN2007bi    & 0.128     & (5,7)  & yes   & 0     & - \\
    SN2010gx    & 0.230     & (5,7)  & yes   & 0     & - \\
    SN2011ke    & 0.143     & (5,7)  & yes   & 0     & - \\
    SN2012il    & 0.175     & (5,7)  & yes   & 5     & 0.8, 1.4, 1.8, 2.2, 3.0 \\
    SN2015bn    & 0.114     & (12)   & yes   & 0     & - \\
    \hline
    \end{tabular}
    \tablefoot{$^a$ Reference to SLSN host study (with \textit{HST}, if available) used here for the host identification, or to SLSN study (if no host study is available): (1) \cite{AngusDES}, (2) \cite{2018ApJ...860..100D}, (3) \cite{2017ApJ...835...58V}, (4) Schulze et al. (in  prep.), (5) \cite{Lunnan2015a}, (6) \cite{Perley2016a}, (7) \cite{Angus2016a}, (8) \cite{CikotaSLSNhosts}, (9) \cite{Thoene2015a}, (10) \cite{Leloudas2012a}, (11) \cite{SUSHIESII}, (12) \cite{2018ApJ...866L..24N}. $^b$ Image completeness: this column describes whether the image depth allows the detection of a companion four times fainter than the host (the limit for our definition of a major merger). $^c$  Number of major companions within 5 kpc.}  
\end{table*}

\begin{table*}[tb]
\caption[Used sample]{Final sample of GRB host galaxies used in our analysis.}
\label{tab:UsedGRBSample}
    \centering
    \begin{tabular}{lrrrr}
    \hline\hline
    Host galaxy & Redshift & N      & Companion distance \\
                &                 &    &    (kpc)\\
    \hline
    GRB970228   & 0.685 & 0     & - \\
    GRB970508   & 0.840 & 0     & - \\
    GRB980613   & 1.100 & 1     & 3.2 \\
    GRB980703   & 0.970 & 0     & - \\
    GRB990123   & 1.600 & 0     & - \\
    GRB990506   & 1.300 & 0     & - \\
    GRB990510   & 1.620 & 0     & - \\
    GRB990705   & 0.860 & 2     & 4.3, 4.8 \\
    GRB990712   & 0.430 & 0     & - \\
    GRB991208   & 0.710 & 0     & - \\
    GRB991216   & 1.020 & 1     & 3.5 \\
    GRB000418   & 1.120 & 0     & - \\
    GRB010222   & 1.470 & 0     & - \\
    GRB010921   & 0.450 & 0     & - \\
    GRB011121   & 0.360 & 0     & - \\
    GRB020405   & 0.690 & 0     & - \\
    GRB020813   & 1.250 & 0     & - \\
    GRB020903   & 0.250 & 1     & 3.6 \\
    GRB021211   & 1.020 & 0     & - \\
    GRB030329   & 0.170 & 0     & - \\
    GRB040924   & 0.859 & 0     & - \\
    GRB041006   & 0.716 & 0     & - \\
    GRB050525A  & 0.606 & 0     & - \\
    GRB051022   & 0.800 & 1     & 3.4 \\
    GRB060729   & 0.540 & 0     & - \\
    GRB070714B  & 0.920 & 0     & - \\
    GRB090113   & 1.749 & 0     & - \\
    GRB100316D  & 0.059 & 5     & 0.6, 1.7, 2.6, 2.9, 3.2 \\
    GRB100615A  & 1.398 & 0     & - \\
    GRB120711A  & 1.405 & 0     & - \\
    GRB130427A  & 0.340 & 2     & 1.7, 3.6 \\
    GRB150314A  & 1.758 & 1     & 2.6 \\
    \hline
    \end{tabular}
\end{table*}

\section{Discussion}
\label{sec:discussion}

Up to this point in the paper, we have shown how SLSN I hosts often make up a part of interacting systems. Here, we discuss the implications of our finding as well as some potential caveats that can influence our interpretation.

\subsection{Implications}

The fact that SLSN I hosts often have companion galaxies at a short separation (within their virial radius) can be related to the production of these rare explosions. Galaxy interaction can boost star formation \citep{Schweizer1986,Barnes1991} and the most natural explanation for the enhanced SLSN I production in these environments is that SLSNe I are also related to recent star formation. This is something that was also proposed by \cite{Leloudas2015a}, where it was argued that SLSNe I might represent the very first explosions following an intense star formation episode, appearing even earlier than GRBs. The fact that we find weaker evidence for interaction among GRB hosts (fewer interacting hosts and fewer major companions, on average, found at larger distances) is consistent with this proposal. We stress, however, that this difference is not statistically significant and is subject to limitations on the sample sizes at comparable redshifts.

Other studies have focused on metallicity as the primary physical factor behind SLSNe I \citep[][]{Lunnan2014a,Perley2016a,Chen2016a}. 
The rationale behind this connection is that metallicity is intimately linked to mass loss and that stars evolving at low metallicities retain more of their angular momentum. 
High angular momentum is an important ingredient for all models that invoke a central engine for a SLSN I \citep[e.g.][]{Kasen2010a,2013ApJ...772...30D}. 
To reconcile the H-free nature of SLSNe I with rapid rotation, however, without the loss of the outer envelope (and significant loss of angular momentum), a homogeneous evolution of the star seems necessary (or spin-up via binary interaction). 
This evolutionary path has also been extensively discussed for GRBs \citep[e.g.][]{2005A&A...443..643Y}. 

There is no doubt that metallicity is a critical factor in SLSN production, as we have previously argued in earlier publications \citep{Leloudas2015a,SUSHIESII}. In particular, in \cite{SUSHIESII}, we derived the most robust limit for the host metallicity of SLSNe I and we showed that the production of SLSNe I is severely stifled above $12+\mathrm{log}(\mathrm{O/H}) > 8.4$. 
However, metallicity alone cannot explain the preference for SLSNe to occur in interacting galaxies. Interaction is strongly pointing towards recent star formation being an equally important (if not defining) factor. 

In this context, metallicity may be a 'prerequisite'\textit{} for SLSNe~I to occur.
If this was not true, we would find more SLSNe~I in massive and metal-rich galaxies (where most of the star formation, including recent star formation, takes place). 
But being a 'prerequisite'\textit{} does not mean that metallicity is the 'cause' behind SLSNe. 
Making this connection a strict causality argument would require adopting the evolutionary path above, which is, of course, plausible but is not without problems. Engine-driven models are not the only viable models for SLSNe. In particular, interaction models, probably combined with some variant of the pulsational pair-instability mechanism \citep{Woosley2007a}, present viable contestants \citep{2018SSRv..214...59M}. Pulsational pair instability also favours, albeit not exclusively, low-metallicity progenitors  \citep{2017ApJ...836..244W}.

If recent star formation plays a causal role behind SLSN production, this must mean that the progenitors of SLSNe I are massive stars that were formed during a recent episode of star formation (which may due to interaction or another mechanism). 
The evolution of these stars in a very metal-poor environment (a prerequisite for SLSNe) is not well understood and it is certainly a key in the production of SLSNe. 
In particular, (episodic) mass loss, the role of binary evolution, and even the possibility of a modified initial mass function in low metallicity starbursts need to be investigated.

Our conclusion that recent star formation and young progenitor age  is an important factor for SLSN production is independent of the exact mechanism behind SLSNe I (engine-driven, interaction-driven or hybrid) but it needs to be taken into account in every model.

\subsection{Potential caveats}

The main limitation of our study is the absence of a control sample of dwarf galaxies, matched in redshift and stellar mass, in order to determine whether the increased number of companions is a property of SLSN hosts or an overall property of dwarf galaxies in general. Unfortunately such a control sample does not exist. Literature measurements of merger fractions are typically conducted on the most luminous or massive galaxies \citep[][and references therein]{Man2016}. The best available study of dwarf galaxy pairs is TiNy Titans \citep[TNT;][]{Stierwalt2015}, which has indeed shown that star formation rate is boosted in dwarf pairs compared to isolated dwarf galaxies. Nevertheless, the observational TNT sample of \cite{Stierwalt2015}, originating from the SDSS, is not suitable to determine the companion fraction of dwarf galaxies (due to the selection methods and biases). An estimate for the companion fraction is instead provided by theoretical calculations and comparison with cosmological simulations \citep{2018MNRAS.480.3376B}. In this study, the authors estimate that the mean number of observed companions per dwarf galaxy should vary between 0.04--0.06 depending on the survey sensitivity limits and correcting for completeness. There are a number of reasons that preclude us from a meaningful comparison with this result. 

Firstly, this study targets a very different (low) redshift interval ($0.01 < z < 0.03$). Secondly, the sample definition criteria are very different than ours: companions are searched for at projected distances $<150$ kpc (a much larger area) but, at the same time, at angular separations $>55$\arcsec to mitigate the effect of fiber confusion in SDSS. Therefore, the very small separations, such as the ones we are studying, are missed. 
Nevertheless, if we were to ignore these differences, we observe that at face value, the companion fractions we are finding for SLSN hosts are ten times larger than the estimates of \cite{2018MNRAS.480.3376B} for dwarf galaxies at low redshift.
This is of course encouraging but without this control sample we cannot definitively answer whether SLSNe hosts are indeed more interacting than other dwarf galaxies.

Another factor that needs to be taken into consideration is the wavelength range for this study.
The \textit{HST} images probe the rest-frame UV. At this wavelength, the galaxies could look more clumpy than in the optical due to the fact they are probing distinct star forming regions.
In addition, UV is more prone to dust extinction but previous studies in the optical have not found any significant evidence for dust in SLSN hosts \citep{Lunnan2014a,Leloudas2015a,Perley2016a}, so we do not think that this can be a driving effect here.
However, as we explained earlier, the sample of SLSN hosts with rest-frame \textit{HST} optical images is very small and such a study would not be currently possible. 
Rest-frame UV imaging was commonly used to identify mergers among high-redshift galaxies before wide-field NIR imaging became available with \textit{HST}/WFC3 \citep[see e.g.][]{Lotz2006,Conselice2008,Conselice2009}. 
In addition, we selected our GRB host comparison sample to also probe rest-frame UV/blue wavelengths.

Redshift evolution and resolution effects have already been discussed in the previous Section.
Although resolution is important for separating a complex structure in multiple components and it favours the detection of more companions at lower redshift, we have shown that this does not shape the high companion fraction among SLSN hosts or the comparison results at least with the Monte Carlo-generated comparison samples. The same is true for all redshift bins that have been studied. 
The comparison with GRB hosts is more sensitive to this effect but the results do not change if we remove the lowest redshift events ($z < 0.1$) from both samples. 
Additionally, even if we have focused our study to a sub-sample of SLSN hosts complete to major mergers ($N = 42$) our conclusions do not significantly change if we use a slightly larger sample ($N = 49$).

Finally, we examine the nature of the companion candidates detected by \texttt{SExtractor} 
and discuss whether they could be galaxies at  different redshifts (chance superposition), foreground stars, or  not  astrophysical at all (cosmic rays). 
The resolution for these problems stems from our choice of comparing with objects from the same images. In all cases, the fact is that the number density of the detected objects remains higher near the SLSN host, irrespective of the nature of the detected objects and it is hard to imagine a bias that would make cosmic rays or foreground stars to be preferentially located near the target of the observation. So even if such biases exist, they cannot be responsible for shaping this result.

Furthermore, the following considerations apply. Pixels affected by cosmic rays are in principle mitigated during the \textit{HST} data processing. As mentioned earlier, however, we had a few cases where this was done inefficiently, leading us to either use a custom reduction or reject the image from our sample. In any case, we inspected the images visually and we determined that only a few are candidates for cosmic ray contamination. Most importantly, we  inspected the regions around the SLSN host and we determined that none of the detections reported as companions in Table~\ref{tab:UsedSample} should be due to cosmic rays. Therefore, cosmic ray contamination would only increase the number density of detections for the regions of the image used for the Monte Carlo-generated comparison samples (\textit{random object, random coordinates}). This effect, if it were corrected for, would only increase the difference between the SLSN host sample and the comparison samples, making the difference more significant than what is presented here. 

This is similar to the case of foreground stars. Certainly, there are such objects in the image and they are indeed used in the \textit{random coordinates}, and even in the \textit{random objects} test, provided that their brightness is similar to the SLSN host. Visual inspection suggests that none of the objects reported as companions in Table~\ref{tab:UsedSample} are star-like. Therefore, the presence of foreground stars would only affect the comparison distributions. In this case, the real difference in the object number density between the SLSN host and the comparison samples would be even larger. We note that we have tried to remove the foreground stars in an automated way, using the star-galaxy separation classifier \texttt{CLASS\_STAR} in \texttt{SExtractor}, but our efforts gave ambiguous results. We concluded that the removal of stars, and not other objects would require a threshold for which only very few would be removed; most stars, except for extreme cases had the same \texttt{CLASS\_STAR} value as other detections, so a \texttt{CLASS\_STAR} threshold would result in many false positives.
As this only affects the exact numbers in the comparison samples, we decided that it was beyond the scope of our paper to explore more sophisticated methods. 

Solving the issue of chance superposition is difficult with the present sample. This would require obtaining the redshifts of all SLSN companions, but this would require substantial spectroscopic time on 8m-class telescopes. Using photometric redshifts is also not possible as most of these fields only have one broad-band \textit{HST} image available. However, at the projected separations that we are looking at (d $<$ 5~kpc), the probability of chance superposition is small: by estimating the surface density of sources of different magnitudes \citep{2002AJ....123.1111B,2010ApJ...722.1946B}, we estimate that only 10\% of our assigned companions have a probability of chance alignment that is $>$10\%.
Ultimately, what matters here is the comparison with our Monte Carlo-generated samples, which has shown that SLSN hosts are found in areas with larger object number density than other areas in the image and there is no reason to think that this would systematically happen due to chance superposition. 
In summary, we are not able to exclude a chance superposition for any individual object that appears as a companion in Table~\ref{tab:UsedSample}. Statistically, however, this should not affect our result.  

We note that for many observing programs in our sample (e.g. 13022, 13326, 15140), the SLSN was deliberately placed in the corner of the detector (including the example case in Fig.~\ref{fig:MapExample}) to mitigate charge transfer efficiency effects (Ragnhild Lunnan, private communication). The sky object density does not depend on the exact location of the target in the detector, so this does not affect our results.  

\section{Conclusion}    
\label{sec:conc}

We conducted a systematic study of SLSN I hosts to determine whether they show signs of interaction with other galaxies. Our final sample consists of 42 SLSN I hosts that have been imaged by the \textit{HST} in the rest-frame UV, over a broad redshift range ($0 < z < 2$).
We ran \texttt{SExtractor} on these images and counted the number of sources that have been detected as separate objects within a radius of 5 kpc around the host galaxy. 
In order to compare it with the SLSN I hosts, we constructed a matched sample of 32 GRB host galaxies and we employed two statistical tests based on Monte Carlo methods applied to the same \textit{HST} images. 
The first Monte Carlo test measures the number of companions for other objects in the image, while the second redistributes the image sources in a random way and re-measures the number of the SLSN host companions in every iteration. The Anderson-Darling K-sample test is used to compare between the distributions of companions and their distances for the SLSN hosts and the comparison samples. We deduce the following:
\renewcommand{\labelitemi}{$\bullet$}
 \begin{itemize}
     \item The SLSN I host galaxies are often part of interacting systems, with $\sim$50\% having at least one major companion within 5~kpc.
     \item The average number of companions for the SLSN host galaxies is $0.70^{+0.19}_{-0.14}$ within 5 kpc for the full redshift range. In comparison, random objects of similar brightness in the same image have on average $0.10^{+0.05}_{-0.05}$ companions. Redistributing the sources in the image result in a average number of $0.06^{+0.02}_{-0.01}$ companions for the SLSN hosts.
     \item Fewer GRB hosts have major companions ($\sim$25\%) and the average number of companions per GRB host is also lower than for SLSNe I: $0.44^{+0.25}_{-0.13}$.
     \item The AD test shows that the difference between the companion distribution of SLSN hosts and those of the Monte Carlo generated comparison samples is statistically significant ($p$-values <$10^{-3}$; Table~\ref{tab:ADtests}) for all redshift bins. 
     \item By constructing the distribution of distances between the SLSN hosts and their companions,
     we find that these are also statistically different from those of our Monte Carlo-generated comparison samples. 
    \item The differences between the SLSN and the GRB distributions are not statistically significant for the number of companions but they are marginally significant for the distance of the companions (SLSN companions are found closer to the host) in the redshift bins $0 < z < 2$ and $0.1 < z < 0.5$ (although the number of GRBs is very low in the latter bin).
 \end{itemize}

We conclude that SLSNe I are often found in interacting environments.
Our interpretation is that SLSNe I are related to a recent burst of star formation, possibly triggered during galaxy interaction. 
Low metallicity is perhaps a stellar evolution prerequisite for SLSN I explosions, as many studies have shown. However, low metallicity alone cannot explain the high interaction fraction of SLSN I host galaxies. This preference  points strongly to recent star formation as an additional critical parameter for SLSN production. This is in line with the large number of starburst galaxies found among SLSN I hosts, as reported in the literature.

\section*{Acknowledgements}

We thank the referee for a constructive report that helped improve this paper.
We are grateful to Andrew Levan for providing us with the reduced images of the \cite{Fruchter2006} GRB host sample.
We wish to thank Ragnhild Lunnan for useful comments on the manuscript, Aleksandar Cikota for sharing reduced images of PTF11hrq, and Daniele Malesani for discussions. 
We acknowledge use of the webpage \url{https://slsn.info/}, maintained by Janet Chen \& Kai Sun, and of a MATLAB procedure for the Anderson-Darling k-sample test by Antonio Trujillo-Ortiz (\url{https://www.mathworks.com/matlabcentral/fileexchange/17451-andarksamtest}).
GL was supported by a research grant (19054) from VILLUM FONDEN.
AM is supported by a Dunlap Fellowship at the Dunlap Institute for Astronomy \& Astrophysics, funded through an endowment established by the David Dunlap family and the University of Toronto. The University of Toronto operates on the traditional land of the Huron-Wendat, the Seneca, and most recently, the Mississaugas of the Credit River; AM is grateful to have the opportunity to work on this land.

\bibliographystyle{aa} 
\bibliography{InteractingSLSNhosts.bib}

\end{document}